\documentclass[journal]{IEEEtran}
\usepackage[colorlinks]{hyperref}
\usepackage{cite}
\usepackage{mathrsfs}
\usepackage{amsthm}
\usepackage{etoolbox} \makeatletter \patchcmd{\@makecaption} {\scshape} {} {} {} \makeatother
\usepackage[cmex10]{amsmath}
\usepackage{float}
\usepackage{mathtools}
 \usepackage{siunitx}
\usepackage{array}
\usepackage{eqparbox}
\usepackage{bm}
\newtheorem{myDef}{Definition}
\usepackage[table,dvipsnames]{xcolor}
\usepackage{multicol,booktabs,tabularx}

\usepackage{cases}
\usepackage{algorithm}
\usepackage{paralist}
\usepackage{algpseudocode}

\usepackage[utf8]{inputenc}
\usepackage[english]{babel}
\usepackage{graphicx}
\usepackage{subfigure}
\usepackage{epstopdf}
\usepackage{epsfig}
\usepackage{amssymb}
\usepackage{array}
\usepackage{multirow}

\usepackage{caption}
\usepackage[numbers,sort&compress]{natbib} % 按引用顺序排序
\begin{document}

\title{Channel Extrapolation for MIMO Systems with the Assistance of Multi-path Information Induced from Channel State Information}
% TODO 重点Time-Aware
\author{Yuan Gao, %~\IEEEmembership{Member,~IEEE},
Xinyi Wu, 
Jiang Jun,
%Zhaohui Yang,~\IEEEmembership{Member,~IEEE},
Zitian Zhang, %~\IEEEmembership{Member,~IEEE},
Zhaohui Yang,
%Shunqing Zhang,
%Jianbo Du,
%Weijie Yuan,
Shugong Xu,~\IEEEmembership{Fellow,~IEEE},\\
Cheng-Xiang Wang,~\IEEEmembership{Fellow,~IEEE},
and Zhu Han,~\IEEEmembership{Fellow,~IEEE}
\thanks{This work was supported by Shanghai Natural Science Foundation under Grant 25ZR1402148. 
(Zitian Zhang and Shugong Xu are the corresponding authors.)
} 
\thanks{Yuan Gao, Xinyi Wu, Jiang Jun 
%and Shunqing Zhang 
are with the School of Communication and Information Engineering, Shanghai University, China, email: gaoyuansie@shu.edu.cn, wu\_xinyi0312@shu.edu.cn, jun\_jiang@shu.edu.cn.} %and shunqing@shu.edu.cn.}
%\thanks{Zhaohui Yang is with the College of Information Science and Electronic Engineering,Zhejiang University, Hangzhou, China,e-mail: yang\_zhaohui@zju.edu.cn.}
\thanks{Zitian Zhang is with School of Information and Electronic Engineering, Zhejiang Gongshang University, Hangzhou, China, e-mail: zitian.zhang@mail.zjgsu.edu.cn.}
\thanks{Zhaohui Yang is with the College of Information Science and Electronic Engineering,Zhejiang University, Hangzhou, China, e-mail: yang\_zhaohui@zju.edu.cn.}
%\thanks{Jianbo Du is with the School of Communication and Information Engineering, Xi'an University of Posts and Telecommunications, Xi'an, China, e-mail: dujianboo@163.com.}
%\thanks{Weijie Yuan is with the Department of Electronic and Electrical Engineering, Southern University of Science and Technology, Shenzhen, China, e-mail: yuanwj@sustech.edu.cn.}
%\thanks{Xiaojun Yuan is with the National Key Laboratory of Wireless Communications, the University of Electronic Science and Technology of China, Chengdu, China, e-mail: xjyuan@uestc.edu.cn.}
\thanks{Shugong Xu is with Xi’an Jiaotong-Liverpool University, Suzhou, China, email: shugong.xu@xjtlu.edu.cn.}
\thanks{Cheng-Xiang Wang is with National Mobile Communications Research Laboratory, School of Information Science and Engineering, Southeast University, Nanjing, China, e-mail: chxwang@seu.edu.cn}
\thanks{Zhu Han is with the Department of Electrical and Computer Engineering at the University of Houston, Houston, USA, and also with the Department of Computer Science and Engineering, Kyung Hee University, Seoul, South Korea, e-mail: hanzhu22@gmail.com.}

%\thanks{Xiaoli Chu is with the Department of Electronic and Electrical Engineering, the University of Sheffield, UK, email: x.chu@sheffield.ac.uk.}
}

\maketitle

% As a general rule, do not put math, special symbols or citations in the abstract or keywords.
\begin{abstract}
Acquiring channel state information (CSI) through traditional methods, such as channel estimation, is increasingly challenging for the emerging sixth generation (6G) mobile networks due to high overhead. To address this issue, channel extrapolation techniques have been proposed to acquire complete CSI from a limited number of known CSIs. To improve extrapolation accuracy, environmental information, such as visual images or radar data, has been utilized, which poses challenges including additional hardware, privacy and multi-modal alignment concerns. To this end, this paper proposes a novel channel extrapolation framework by leveraging environment-related multi-path characteristics induced directly from CSI without integrating additional modalities. Specifically, we propose utilizing the multi-path characteristics in the form of power-delay profile (PDP), which is acquired using a CSI-to-PDP module. CSI-to-PDP module is trained in an AE-based framework by reconstructing the PDPs and constraining the latent low-dimensional features to represent the CSI. We further extract the total power \& power-weighted delay of all the identified paths in PDP as the multi-path information. Building on this, we proposed a MAE architecture trained in a self-supervised manner to perform channel extrapolation. Unlike standard MAE approaches, our method employs separate encoders to extract features from the masked CSI and the multi-path information, which are then fused by a cross-attention module. Extensive simulations demonstrate that this framework improves extrapolation performance by approximately 4–5 dB, with a minor increase in inference time (around 0.1 ms) on an NVIDIA GeForce RTX 4090 GPU. Notably, the most influential multi-path information for high-performance channel extrapolation is the total path power and the power-weighted delay, which encapsulate comprehensive environmental information. Furthermore, our model shows strong generalization capabilities, particularly when only a small portion of the CSI is known, outperforming existing benchmarks.
\end{abstract}


% \fi
\IEEEpeerreviewmaketitle

\section{Introduction}

Channel extrapolation is considered as a key enable for sixth generation (6G) to acquire channel state information (CSI) with reduced overhead \cite{zhang2023ai,gao2026csiextra,wang2022pervasive,wang2025enhanced}. Ultra-high-speed communication scenarios, such as unmanned aerial vehicles (UAVs), vehicle-to-everything (V2X), and high-speed trains (HST), are a key application direction of 6G \cite{gao2024performance,jin2024efficient,gao2025stochastic,du2024secure}. Channel aging poses a significant challenge in these cases due to the rapid shifts in wireless environment \cite{jin2025linformer}, and requires performing conventional CSI acquisition techniques, such as channel estimation, more frequently, contributing to substantial overhead. Time-domain channel extrapolation leverages historical CSI to predict future CSI based on temporal correlations, thereby minimizing the frequency of CSI estimations \cite{jiang2025towards}. For multiple-input multiple-output (MIMO) systems in fourth generation (4G), dedicated pilots are allocated to each transmit (Tx) antenna and receive (Rx) antenna pair to avoid interference for channel estimation. Such channel estimation technique is non-scalable for ultra-massive MIMO in 6G due to excessive overhead. To this end, antenna-domain channel extrapolation is proposed to extrapolate CSI of Tx-Rx pairs of interest by utilizing known CSI Tx-Rx pairs, thereby effectively reducing the overhead for channel acquisition \cite{jin2025context}.

Research on channel extrapolation has evolved from classical statistical methods to sophisticated data-driven approaches. Early efforts relied on autoregressive (AR) models \cite{kim2020massive,  prakash2011effects,yin2020addressing,wu2013optimal,pedrosa2020efficient,wang2024two,peng2017channel,li2022multi}, and Kalman filters \cite{kim2020massive,pedrosa2020efficient,prakash2011effects,yang2018secure} to predict CSI in time-varying channels, leveraging linear assumptions for simplicity and low computational demand. Besides, parametric channel-based methods are based on the assumption that channel characteristics can be described by a specific set of parameters, and channel extrapolation across time, frequency, and spatial domains is performed by estimating variations in these parameters \cite{peng2022novel,ren2014position,    fleury1999channel,sun2022channel, guo2017millimeter,  uehashi2018prediction,shi2023channel,xu2022sparse,xiao2023nonparametric}. The performance of the above approaches highly depends on the alignment between the channel model (assumption) and the practical wireless channel. More recent studies have incorporated machine learning and deep learning techniques \cite{gao2025joint}, for instance, deep learning models like long short-term memory (LSTM) networks have been applied to extrapolate CSI in massive MIMO systems, demonstrating improved accuracy over traditional methods in fading environments \cite{soszka2022fading,zhang2022deeplearning}. These advancements have extended to frequency and spatial domains, with techniques like convolutional neural networks (CNNs) \cite{jiang2023novel,zhou2025low,gong2024lightweight}, graph neural networks (GNNs) \cite{zhang2024learning}, and Transformer \cite{jin2025linformer,wang2024deep,gong2024deep,jin2025context}.

However, artificial intelligence (AI)-enabled channel prediction models are typically optimized for specific scenario, which leads to the challenge of generalization for diverse scenarios. Building on these foundations, environment-assisted channel extrapolation has gained traction by incorporating auxiliary data to refine predictions, including vision \cite{moon2025multi,xin2024novel}, radar \cite{ren2024sensing} and wireless sensing information \cite{he2025sensing}. The work in \cite{moon2025multi} proposed a vision-assisted channel prediction algorithm for millimeter wave (mmWave) massive antenna systems by exploiting both vision sensing and wireless signaling to identify the orientations and positions of mobile devices, and materials of surrounding wireless environments (e.g., walls and ground). \cite{xin2024novel} proposed novel unmanned aerial vehicle (UAV)-to-ground channel prediction method in the three dimensional (3D) airspace by integrating UAV-captured images, location data of transmitters and receivers, and communication settings. The channel prediction accuracy is enhanced by utilized the additional vision information of the environment. \cite{ren2024sensing} proposed to utilize the wireless channel sparsity and multi-path effect acquired via radar sensing by detecting the scatterers in the environment. Similarly, \cite{he2025sensing} proposed to enhance the channel prediction performance by jointly utilizing the communication and sensing CSI. Hybrid schemes combining CSI with digital twins or channel knowledge maps (CKMs) is an emerging type of environment information, which is promising to enhance the performance of channel extrapolation \cite{wu2023environment,qi2025novel}. These environment-assisted strategies often outperform pure CSI-based methods by leveraging contextual information for more accurate long-term predictions.

Despite these progresses, existing research on channel extrapolation faces several limitations that hinder its deployment in practical 6G systems. \begin{itemize}
\item Integrating external sensors for environmental assistance often requires dedicated equipment beyond standard wireless transceivers, leading to higher deployment and maintenance expenses. For instance, vision-aided systems might necessitate high-resolution cameras or LiDAR units at base stations (BSs) or user devices, which add to the overall system complexity and cost. \item Additional, the use of visual or sensory data in environment-assisted methods raises dramatic privacy concerns, as these modalities can inadvertently capture sensitive information about individuals or surroundings. Deploying cameras for blockage prediction or beam alignment, for example, might record personal activities in public spaces, leading to regulatory hurdles and user distrust. \item Moreover, aligning data from diverse modalities, such as synchronizing visual images with CSI measurements, presents significant technical difficulties due to differences in sampling rates, spatial resolutions, and temporal dynamics. In vision-assisted beamforming, for example, mismatches in coordinate systems between camera views and radio propagation paths can lead to inaccurate predictions, requiring complex calibration processes. High-mobility scenarios exacerbate such problem, as rapid changes in user positions can desynchronize multimodal inputs, resulting in fusion errors and degraded extrapolation performance. 
\end{itemize}

This paper introduces a novel framework for channel extrapolation that addresses challenges in aligning multi-modal data, reducing hardware costs, and protecting user privacy, by leveraging environment-related multi-path information derived directly from channel state information (CSI). Our approach uses the power delay profile (PDP) to capture key propagation features, such as power and delay, which reflect the surrounding environment. The proposed framework integrates a CSI-to-PDP conversion module with a channel extrapolation module. The key contributions of this work are outlined below:

\begin{itemize}
\item 
We develop a CSI-to-PDP module that extracts the PDP from CSI using a decoder trained in an auto-encoder (AE) framework. The AE is designed to reconstruct the PDP while ensuring the encoder’s output closely represents the CSI, which is different from the conventional AE framework that takes latent low-dimensional representation as the output of encoder. The training process minimizes the summation of PDP reconstruction errors and CSI representation errors. To assist subsequent channel extrapolation, we proposed to extract the total power \& power-weighted delay of all the identified paths in PDP.   

\item We propose a masked auto-encoder (MAE) architecture for channel extrapolation, trained in a self-supervised manner. Unlike traditional MAE models, our approach uses separate encoders to extract features from masked CSI and the total power and power-weighted delay. These features are combined using a cross-attention-based fusion module, which captures long-range relationships between them. Specifically, the total power and power-weighted delay serve as the query, while CSI features act as the key and value, enhancing extrapolation accuracy.

\item Extensive simulation results show that, with the assistance of the total power \& power-weighted delay, our framework improves extrapolation accuracy by approximately 4–5 dB at the expense of a slight increase in inference time (around 0.1 ms) on an NVIDIA GeForce RTX 4090 GPU. Our analysis indicates that the total power \& power-weighted delay are most influential for performance enhancement of channel extrapolation. The proposed feature fusion module outperforms feature concatenation and baseline cross-attention scheme. Additionally, the proposed model demonstrates strong generalization ability in different frequency band, especially when only a small portion of the CSI is available, outperforming existing methods on channel extrapolation. 
\end{itemize}

The remaining paper is organized as follows. We formulate the problem of channel extrapolation in Section \ref{Sec:system_model}. In Section \ref{Sec:proposed_model}, we elaborate the proposed channel extrapolation framework, especially the AE-based model that acquires PDP directly from CSI, and the MAE-based channel extrapolation framework. Extensive simulation results with in-depth analysis are presented in Section \ref{Sec:simulations}. Finally, Section \ref{Sec:conclusions} ends this paper with conclusions.
\section{Problem formulation of Channel Extrapolation}
\label{Sec:system_model}
\subsection{Channel Model}

\begin{figure*}[t]  
\centering
\includegraphics[width=0.9\textwidth]{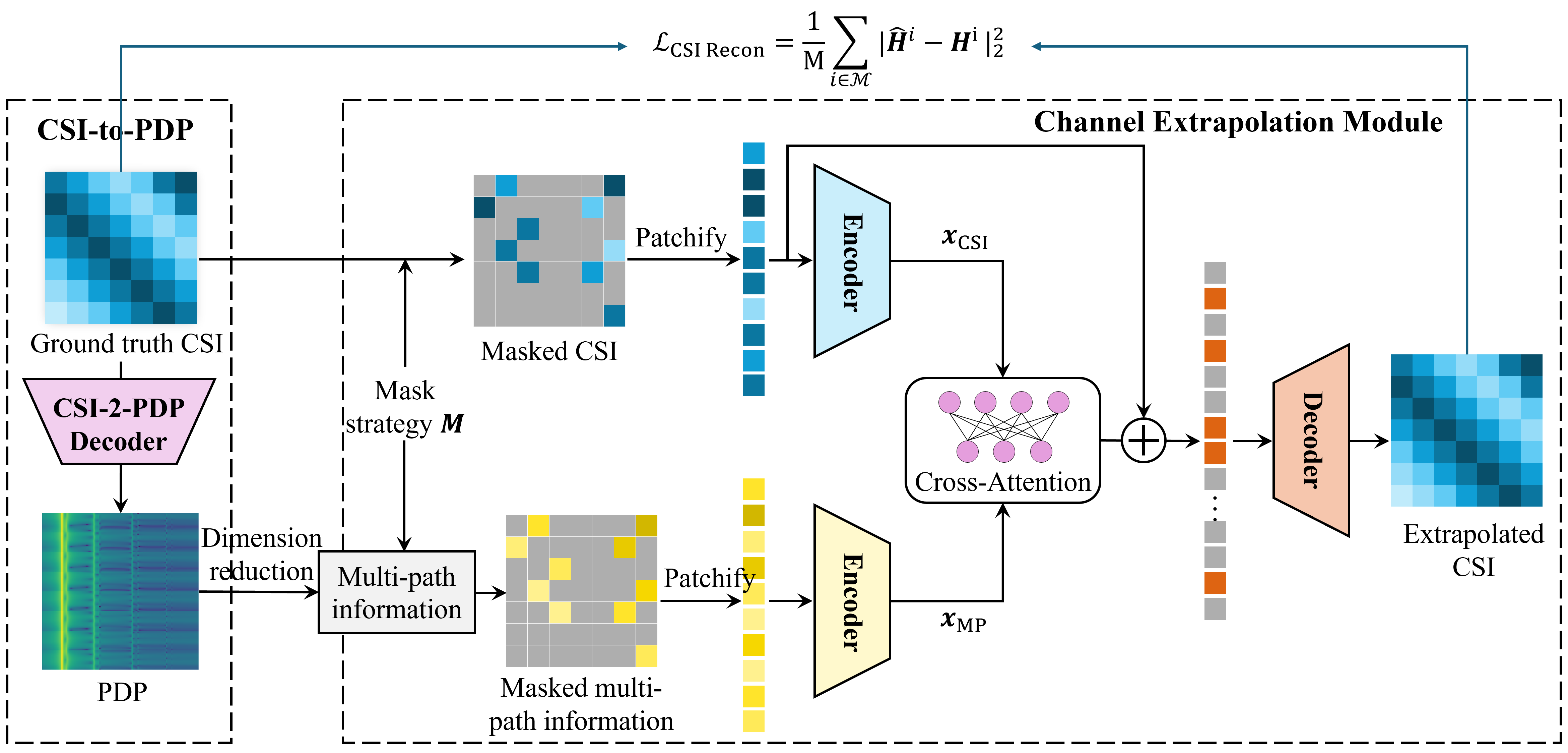} 
\caption{Overal architecture of the proposed model, which consists of a CSI-to-PDP and channel extrapolation modules. The CSI-to-PDP module infers PDP using CSI via a CSI-to-PDP decoder, which is training in a auto-encoder (AE) manner illustrated in Fig. \ref{fig:CSI-to-PDP}. The channel extrapolation module is developed based on a masked auto-encoder (MAE), where the features of masked CSI and PDP indicators are extracted separately using dedicated encoders and fused subsequently. The fused features are finally utilized to reconstruct the complete CSI using a decoder. }
\label{fig:architecture} 
\end{figure*}

\label{sec:system_modeling}
We consider a multi-antenna wireless communications, where the BS and user equipment (UE) are equipped with $K$ transmit and $M$ receive antennas, respectively. The CSI between the BS and UE is denoted by the channel matrix $\boldsymbol{H} \in \mathbb{C}^{M \times K}$ as:

\begin{equation}
  \boldsymbol{H} = \left(
    \begin{array}{ccc}
      h_{11} & \cdots & h_{1K}\\
      \vdots & \ddots & \vdots\\
      h_{M1} & \cdots & h_{MK}\\
    \end{array}
  \right),
\end{equation}
where $h_{mk}$ is the CSI between the $m$-th antenna at BS and the $k$-th antenna at UE.

To capture the wireless propagation characteristics, such as the prorogation attenuation, multi-path effect, etc, we consider the following channel model:
\begin{equation}
  h_{mk}=\sum_{l_{mk}=1}^{L_{mk}}\alpha_{l_{mk}} e^{j \phi_{l_{mk}}} \mathbf{a}_r(\theta_{l_{mk}}^r, \phi_{l_{mk}}^r) \mathbf{a}_t^H(\theta_{l_{mk}}^t, \phi_{l_{mk}}^t),
\end{equation}
where $L_{mk}$ is the number of propagation paths between the $m$-th antenna at BS and the $k$-th antenna at UE. $\alpha_{l_{mk}}$ and $\phi_{l_{mk}}$ are the amplitude and phase of the propagation path $L_{mk}$, respectively. $\mathbf{a}_r(\theta_{l_{mk}}^r, \phi_{l_{mk}}^r)$ is the receive array response vector with respect to the azimuth and elevation angles of arrival (AoA) $\theta_{l_{mk}}^r$ and $\phi_{l_{mk}}^r$. $\mathbf{a}_t(\theta_{l_{mk}}^t, \phi_{l_{mk}}^t)$ is the transmit array response vector with respect to the azimuth and elevation angles of departure (AoD) $\theta_{l_{mk}}^t$ and $\phi_{l_{mk}}^t$. $(\cdot)^H$ is the Hermitian transpose.

As the adjacent antennas are placed within a distance at level of wireless signal wavelength, the difference between amplitude $\alpha_{l_{mk}}$,  phase $\phi_{l_{mk}}$, azimuth AoA $\theta_{l_{mk}}^r$, elevation AoA $\phi_{l_{mk}}^r$, azimuth AoD $\theta_{l_{mk}}^t$ and elevation AoD $\phi_{l_{mk}}^t$ of the propagation path between the $m$-th antenna at BS and the $k$-th antenna at UE are small. However, with the increase of the distance between antennas, spatial nonstationarity becomes significant, for example, according to the research on channel measurement for MIMO systems, not all the Tx antenna and Rx antenna pairs share the same propagation paths \cite{li20183d}. Conventional, CSI for MIMO systems is acquired by channel estimation via pilot signal. To improve the performance of channel estimation and avoid pilot contamination, orthogonal pilot is allocated to each Tx-Rx antenna pair. For massive MIMO systems, the overhead of pilots increase dramatically and becomes significant. 

\subsection{Problem Formulation}
In the existing research \cite{yin2020addressing,wu2013optimal,wang2024two,peng2017channel,li2022multi,kim2020massive,pedrosa2020efficient,prakash2011effects,yang2018secure,peng2022novel,ren2014position,    fleury1999channel,sun2022channel, guo2017millimeter,   uehashi2018prediction,shi2023channel,xu2022sparse,xiao2023nonparametric, gao2025joint, gao2025ssnet,gao2025enabling,soszka2022fading,zhang2022deeplearning, jiang2023novel,zhou2025low,gong2024lightweight, zhang2024learning, jin2025linformer,wang2024deep,gong2024deep,jin2025context}, channel extrapolation has been proposed to extrapolate the complete channel matrix $\boldsymbol{H}$ using a limited number of $\{h_{mk}\}$, which is expressed mathematically as: 
\begin{equation}
\widehat{\boldsymbol{H}}=f_\theta(\{h_{mk}\}),
\end{equation}where $\widehat{\boldsymbol{H}}$ is the extrapolated channel matrix. $f_\theta(\{h_{mk}\})$ is the function that extrapolates the complete channel matrix using partial CSI $\{h_{mk}\}$. 

To further enhance the performance of channel extrapolation, we propose to utilize environment information to assist channel extrapolation. Specifically, the environment directly affect the multi-path characteristic of the wireless signal, including propagation, reflection, diffraction, etc. PDP $\mathbf{P}_{mk}$ indicates the received power of multiple paths against the delay between the $m$-th antenna at BS and the $k$-th antenna at UE, which quantifies the multi-path characteristic of the wireless signal as:
\begin{equation}
\mathbf{P}_{mk}=\{{p}^{i}_{mk},\tau^i_{mk}\}, i\in \mathbb{S}_{mk},
\end{equation}where $\mathbb{S}_{mk}$ is the set of time steps of PDP between the $m$-th antenna at BS and the $k$-th antenna at UE. ${p}^{i}_{mk}$ and $\tau^i_{mk}$ are the power and delay of at the $i$-th path, respectively. Different from existing research that integrates environment information in additional modalities \cite{moon2025multi,xin2024novel,ren2024sensing,he2025sensing,wu2023environment}, we propose to acquire PDP $\{\mathbf{P}_{mk}\}$ using corresponding CSI $\{h_{mk}\}$ as: 
\begin{equation}
\{\mathbf{P}_{mk}\}=f_\text{{CSI-to-PDP}}(\{h_{mk}\})\,
\end{equation}where $f_\text{{CSI-to-PDP}}$ is the mapping between CSI and PDP. 

\begin{myDef} \textbf{Multi-path information}: as the dimension of PDP is enormously high, a dimensionality reduction function $f_\text{DR}$ is further performed to obtain a low-dimensional PDP feature, i.e., multi-path information $\{\mathbf{P}'_{mk}\}$ as:
\begin{equation}
\{\mathbf{P}'_{mk}\}=f_\text{DR}(f_\text{{CSI-to-PDP}}(\{h_{mk}\})).
\end{equation}
\end{myDef}

$\{\mathbf{P}'_{mk}\}$ is further utilized for channel extrapolation as:
\begin{equation}
\widehat{\boldsymbol{H}}'=f'_\theta\left(\{h_{mk}\},\{\mathbf{P}'_{mk}\}\right),
\end{equation}where $\widehat{\boldsymbol{H}}'$ is the extrapolated channel metric and $f'_\theta$ is the channel extrapolation function. The objective is to find $f'_\theta$ to minimize the mean squared error (MSE) between the extrapolated CSI $\widehat{\boldsymbol{H}}'$ and the ground truth CSI $\boldsymbol{H}$, which is formulated as follows:
\begin{equation}
     \min_{f'_\theta}(|\widehat{\boldsymbol{H}'}-\boldsymbol{H}|^2).
\end{equation}

\section{Proposed model} 
\label{Sec:proposed_model}
To effective utilize the multi-path information, we proposed an MAE-based model, which is illustrated in Fig. \ref{fig:architecture}. The proposed model consists of a CSI-to-PDP and channel extrapolation modules. The CSI-to-PDP module infers PDP using CSI via a CSI-to-PDP decoder, which is training in an AE manner illustrated in Fig. \ref{fig:CSI-to-PDP}. The channel extrapolation module is developed based on a MAE, where the features of masked CSI and PDP indicators are extracted separately using dedicated encoders and fused subsequently. The fused features are finally utilized to reconstruct the complete CSI using a decoder. The detail of the proposed model is elaborated in this section.

%\begin{figure*}[t]  
%\centering
%\includegraphics[width=\textwidth]{architecture.png} 
%\caption{Architecture of the proposed model}
%\label{fig:model}  % 标签
%\end{figure*}

%\subsection{CSI-to-PDP}
%\begin{figure}[t]  
%\centering
%\includegraphics[width=0.5\textwidth]{CSI-2-PDP.png} 
%\caption{}
%\label{fig:CSI-to-PDP}  % 标签
%\end{figure}
\subsection{CSI-to-PDP}

\begin{figure}[t]  
\centering
\includegraphics[width=0.5\textwidth]{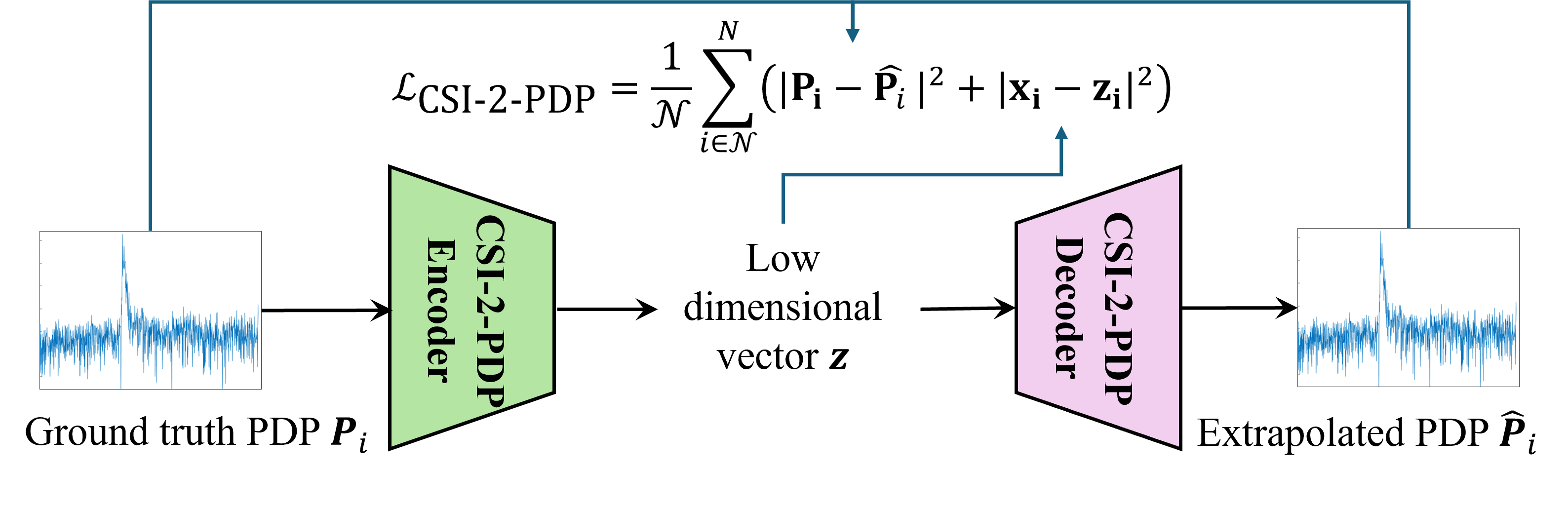} 
\caption{Training procedure of the CSI-to-PDP in an auto-encoder (AE) manner. To enable the mapping between CSI and PDP, the output of the CSI-to-PDP encoder is forced to be CSI by using the summation of reconstruction error of CSI and PDP.}
\label{fig:CSI-to-PDP} 
\end{figure}

As illustrated in Fig. \ref{fig:CSI-to-PDP}, we propose an AE-based framework \cite{wang2016auto,zhai2018autoencoder} to extrapolate the PDP using CSI. AE models generally leverage unsupervised learning to derive a compact representation from input data through two main components: the encoder and the decoder. The encoder compresses the input data into a lower-dimensional representation, while the decoder reconstructs the original data from this compressed form. 

To exploit the intrinsic ability of AE to learn disentangled, probabilistic latent spaces that not only compress high-dimensional data but also enforce a structured manifold amenable to generative extrapolation, we change the conventional AE to suit for CSI-to-PDP extrapolation. Particularly, the PDP serves dually as both the input to the encoder and the target output of the decoder, with the latent representation explicitly modeled as the CSI. The improved framework introduces two critical enhancements: First, by explicitly modeling CSI as the latent variable of PDP, the latent representation acquires well-defined physical significance, which facilitates system debugging and theoretical analysis. Second, the proposed joint loss function (combining CSI reconstruction and PDP reconstruction) simultaneously constrains both the latent space and output space, enabling better capture of the physical correlations between PDP and CSI. This dual-constraint mechanism ensures the learned latent space preserves essential physical characteristics while maintaining high-fidelity PDP reconstruction capability.

Let $\mathbf{P}$ represent the PDP and $\mathbf{C}$ be the corresponding CSI. The encoder is defined as a nonlinear mapping $f_E(\mathbf{P}; \theta_E)$, where $\theta_E$ is the parameters of the encoder. This mapping converts the PDP $\mathbf{P}$ into a lower-dimensional vector $\mathbf{z} = f_E(\mathbf{P}; \theta_E)$. Ideally, this representation $\mathbf{z}$ should capture essential features of the input and closely resemble the target CSI $\mathbf{C}$, i.e., $\mathbf{z} \approx \mathbf{C}$.

The decoder reconstructs the original PDP $\mathbf{P}$ using the output $\mathbf{z}$ of the encoder. Similarly, the process is denoted by a nonlinear mapping $f_D(\mathbf{z}; \theta_D)$, where $\theta_D$ is the parameters of the decoder. The reconstructed PDP $\hat{\mathbf{P}}$ should be nearly identical to the original input PDP $\mathbf{P}$, expressed as:

\begin{equation}
\hat{\mathbf{P}} = f_D(f_E(\mathbf{P}; \theta_E); \theta_D) \approx \mathbf{P}.
\end{equation}

The training objective is to minimize both the reconstruction error and the difference between $\mathbf{z}_i$ and the true CSI $\mathbf{x}_i$, as follows
%\begin{equation}
\begin{equation}
\mathcal{L}_{\text{CSI-to-PDP}} = \frac{1}{N } \sum_{i\in \mathcal{N}}^{N} \left( \left\| \mathbf{P}_i - \hat{\mathbf{P}}_i \right\|^2 + \left\| \mathbf{x}_i - \mathbf{z}_i \right\|^2 \right),
\end{equation}where $\mathcal{N}$ and $N$ are the dataset of PDP-CSI pairs and the number of samples in $\mathcal{N}$, respectively.

\subsection{Channel Extrapolation}

In the channel extrapolation module, we propose an enhanced architecture based on the MAE \cite{he2022masked}, designed to reconstruct complete CSI from partially observed CSI. The model preserves the core principles of MAE, including a high-ratio masking strategy and an asymmetric encoder-decoder framework under a self-supervised paradigm, while introducing dual-branch processing to handle CSI and multi-path information extracted from the PDP. After applying random masking, the visible subset is fed into the encoder for joint feature extraction and cross-modal fusion. The decoder then reconstructs the full CSI  from the encoded latent representations. The loss function is MSE, expressed as:

\begin{equation}
\mathcal{L}_{\text{CSI Recon}} = \frac{1}{\mathcal{M} } \sum_{i \in \mathcal{M}}^{M} \left\| \boldsymbol{H}_i-\widehat{\boldsymbol{H}}_i \right\|
^2,
\end{equation} where $\mathcal{M}$ is the set of masked CSI indices.

\subsubsection{Pre-processing and Mask Strategy}

The input CSI can be expressed as $\mathbf{X}_\text{CSI}\in\mathbb{R}^{B \times 2 \times M \times K}$, where $B$ represents the batch size, 2 represents the real and imaginary part, $K$ and $M$ represent the number of transmit and receive antennas respectively. 

Instead of using the complete PDP as input, we extract representative features to reduce computational complexity. Specifically, we set a power threshold ${p}^\text{threshold}_{mk}=\frac{1}{3}{p}^\text{max}_{mk}$ (${p}^\text{max}_{mk}$ is the maximum power of all paths) and retain only paths with power above this threshold, regarding them as effective paths \cite{vales2020using,ai2015power}. We then extract two key features: the sum of the power ${p}^\text{total}_{mk}$, which reflects the total energy of the multi-path signal, and the power-weighted average delay ${\tau}^\text{w}_{mk}$, which characterizes the temporal dispersion of the channel. Mathematically, total power and power-weighted average delay can be respectively expressed as:
\begin{align}
{p}^\text{total}_{mk}=\sum_{{i}\in\mathbb{P}}{p}^{i}_{mk},\label{eq:total_power}\\
\mathbf{\tau}^\text{w}_{mk}=\frac{\sum_{{i}\in\mathbb{P}}{p}^{i}_{mk}\mathbf{\tau}^{i}_{mk}}{\sum_{{i}\in \mathbb{p}}{p}^{i}_{mk}},
\label{eq:weighted_delay}\end{align}where $\mathbb{P}=\{\mathbf{p}_\text{1},\mathbf{p}_\text{2},\mathbf{p}_\text{3},...,\mathbf{p}_\text{n}\}$ is the set of effective paths between the $m$-th antenna at BS and the $k$-th antenna at UE. The ${p}^\text{total}_{mk}$ and $\mathbf{\tau}^\text{w}_{mk}$ constitute the multi-path information $\mathbf{X}_\text{MP}\in\mathbb{R}^{B \times 2 \times M \times K}$, which maintains the same dimensional structure as the $\mathbf{X}_\text{CSI}$ to ensure architectural consistency. Both $\mathbf{X}_\text{CSI}$ and $\mathbf{X}_\text{MP}$ undergo Z-score normalization to standardize the feature distributions and accelerate training convergence, implemented as:
\begin{align}
\mathbf{X}_\text{CSI}^\text{norm} = \frac{X_\text{CSI}-\mu_\text{CSI}}{\sigma_\text{CSI}},\\
\mathbf{X}_\text{MP}^\text{norm} = \frac{X_\text{MP}-\mu_\text{MP}}{\sigma_\text{MP}},
\end{align} where $\mu_\text{CSI}$ and $\sigma_\text{CSI}$ are the mean and standard deviation of CSI respectively, and $\mu_\text{MP}$ and $\sigma_\text{MP}$ are the mean and standard deviation of multi-path information, respectively.

The normalized features are processed through a shared random masking strategy that applies identical masking patterns to both modalities to preserve their spatial correspondence. This strategy begins by generating a uniform random matrix $K\in[0,1]$, where each element represents an independent sampling from a continuous uniform distribution, serving as the basis for determining the retention priority of each position in the sequence. The algorithm then performs an ascending sort operation on the noise values along the sequence dimension to obtain the shuffled indices $\mathbf{Ids}_\text{shuffle}$, where positions with smaller noise values are ranked higher in the preservation order. Based on the predetermined mask ratio $\rho$, the number of retained tokens is $L_\text{keep} = L_\text{total} \times (1 - \rho)$, and the algorithm selects the first $L_\text{keep}$ indices from $\mathbf{Ids}_\text{shuffle}$ to form the preserved position set $\mathbf{Ids}_\text{keep}$. A binary mask matrix $\textbf{M}\in \mathbb{R}^{B \times M\times K}$ is constructed where zeros indicating preserved positions and ones representing masked locations, enabling the calculation of the reconstruction loss during training. The final masked sequences $\mathbf{X}_\text{CSI}^\text{masked}$ and $\mathbf{X}_\text{MP}^\text{masked}$ are obtained by gathering only the preserved tokens from the original sequences, resulting in partial representations of the antenna dimension information for subsequent encoding.

\subsubsection{Encoder}

The model contains two separate parallel encoders to extract relevant information from the unmasked CSI and multipath characteristics, generating latent representations. Let $\mathbf{X}_\text{input}^\text{enc}\in\mathbb{R}^{B\times2\times M\times K}$  represents either $\mathbf{X}_\text{CSI}^\text{masked}$ or  $\mathbf{X}_\text{MP}^\text{masked}$. In each encoder, the ${X}_\text{input}$ is first transformed into serialized token representations through convolution operations:

\begin{equation}
\mathbf{X}_\text{conv} = \text{Conv2D}(\mathbf{X}_\text{input}^\text{enc}).
\end{equation}  

The output $\mathbf{X}_\text{conv}\in \mathbb{R}^{B \times D \times M' \times K'}$, where the kernel size and stride are both equal to the patch size $p\times p$, $M' = M/p$, $K' = K/p$. Then, we use the flatten layer to generate the patch embedding sequence, mapping each spatial patch region to a D-dimensional embedding vector $\mathbf{X}_\text{patch}\in \mathbb{R}^{B \times L \times D}$, where $ L=M' \times K'$ is the number of patches. However, while this operation preserves the local features within each image patch, it completely loses the relative positional information of the patches within the original 2D space. To incorporate the spatial information of the antenna ports, the model element-wise adds the positional encoding $\mathbf{E}_\text{pos}$ to the embedding $\mathbf{X}_\text{patch}$ as:
 
\begin{equation}
\mathbf{X}_\text{pos} = \mathbf{X}_\text{patch} + \mathbf{E}_\text{pos}.
\end{equation}

To ensure that the $\mathbf{E}_\text{pos}$ is modality-invariant for both CSI and PDP inputs, we we adopt the two dimensional (2D) sine-cosine positional encoding that is only determined by the grid location ($i$, $j$) of each patch, where $i$ and $j$ are the row and column indices in the $\sqrt{L}*\sqrt{L}$ patch grid respectively. The $D$-dimensional positional embedding matrix $\mathbf{E}_\text{pos}\in \mathbb{R}^{L \times D}$ is constructed as:

\begin{equation}
\mathbf{E}_\text{pos} = \text{Flatten}(\mathbf{E}_\text{grid}), 
\end{equation} where $\mathbf{E}_\text{grid}\in \mathbb{R}^D$ is computed for each grid position ($i$, $j$) by:

\begin{equation}
\mathbf{E}_\text{grid}(i,j) = \text{Concat}\left( \mathbf{E}_\text{row}(i), \mathbf{E}_\text{col}(j) \right),
\end{equation} with $\mathbf{E}_\text{row},\mathbf{E}_\text{col}\in \mathbb{R}^{{G} \times D/2}$ being the row and column encoding matrices respectively. The sinusoidal components are calculated as:

\begin{align}
\mathbf{E}_\text{row}(i,2k) = \sin(i/\omega_k),\\
\mathbf{E}_\text{row}(i,2k+1) = \cos(i/\omega_k),\\
\mathbf{E}_\text{col}(j,2k) = \sin(j/\omega_k),\\
\mathbf{E}_\text{col}(j,2k+1) = \cos(j/\omega_k),
\end{align} where $\omega_k=10000^{2k/D}$ for $0 \leq k < \lfloor D/4 \rfloor$.

The patch embeddings with positional information are then processed through a stack of identical Transformer blocks, which consists of a multi-head self-attention (MSA) modeul followed by a feed-forward network (FFN), with residual connections and layer normalization applied at each stage. Each block first applies layer normalization to the ${X}_\text{pos}$ to stabilize activations and accelerate convergence, followed by the MSA module computes attention across all patches:

\begin{equation}
\text{MSA}(\mathbf{X}_\text{pos}) = \text{Concat}(\text{head}_1, ..., \text{head}_h)\mathbf{W}^\text{O},
\end{equation} where each head implements scaled dot-product attention as:
\begin{equation}
\text{head}_i = \text{Softmax}\left(\frac{(\mathbf{X}_\text{pos}\mathbf{W}i^\text{Q})(\mathbf{X}_\text{pos}\mathbf{W}i^K)^T}{\sqrt{d_k}}\right)\mathbf{X}_\text{pos}\mathbf{W}i^V,
\end{equation} with learnable projections matrices $\mathbf{W}i^\text{Q}, \mathbf{W}i^\text{K}\in\mathbb{R}^{D \times d_k}$, $ \mathbf{W}i^\text{V}\in\mathbb{R}^{D \times d_v}$ and output projection matrix $\mathbf{W}^O\in\mathbb{R}^{hd_v \times D}$.

The attention output is combined with the original input via a residual connection after stochastic depth regularization:
\begin{equation}
\mathbf{X}_\text{block} = \mathbf{X}_\text{pos} + \text{DropPath}(\text{MSA}(\mathbf{X}_\text{pos})),
\end{equation} followed by another layer normalization operation that prepares features for the subsequent FFN. The FNN enhances feature representation through an expanded hidden dimension by first projecting features to $\mathbb{R}^{4D}$ space with GELU activation using the projextion matrix  $\mathbf{W}_1\in\mathbb{R}^{D \times 4D}$, then projecting back to the original dimension through $\mathbf{W}_2\in\mathbb{R}^{4D \times D}$, with the final output computed as:
\begin{align}
\mathbf{X}_{\text{out}} = \mathbf{X}_\text{block} + \text{DropPath}(\text{FFN}(\mathbf{X}_\text{block})),\\
\text{FFN}(\mathbf{X}_\text{block}) = \text{GELU}(\mathbf{X}_\text{block}\mathbf{W}_1 + \mathbf{b}_1)\mathbf{W}_2 + \mathbf{b}_2.
\end{align}

Both CSI and multipath characteristics map are independently processed through such encoder to compute the latent feature $\mathbf{Z}_\text{CSI}\in\mathbb{R}^{B\times L\times D}$ and $\mathbf{Z}_\text{MP}\in\mathbb{R}^{B\times L\times D}$.

\subsubsection{Feature Fusion}
\begin{figure}[t]  
\centering
\includegraphics[width=0.4\textwidth]{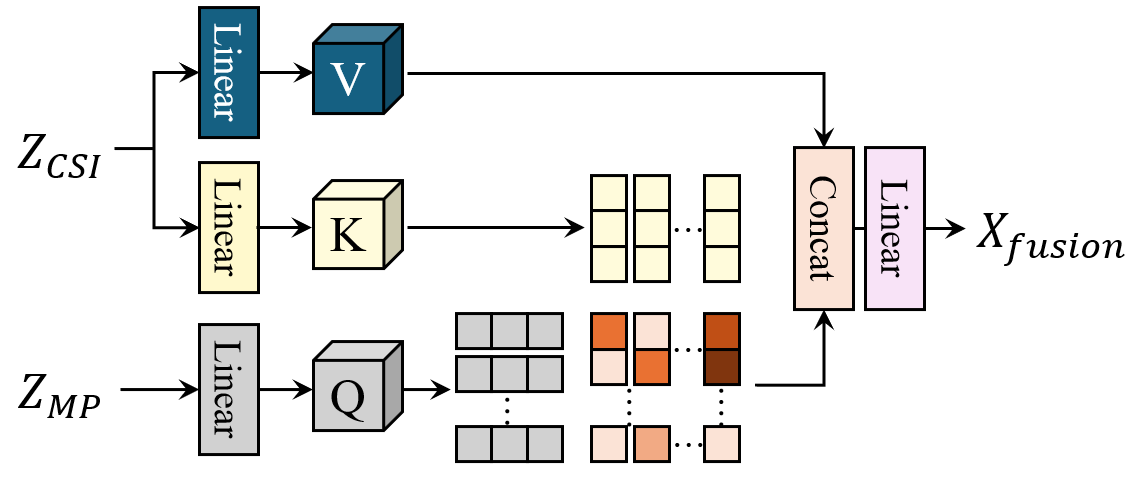} 
\caption{Fusion of the multi-path information and CSI features.}
\label{fig:feature_fusion} 
\end{figure}
After the encoder, we propose the cross-attention mechanism to fuse the latent feature $\mathbf{Z}_\text{CSI}$ and $\mathbf{Z}_\text{MP}$. This establishes interactions between CSI and multipath characteristics domains through query-key-value mapping, enabling the model to focus on the most important information of both modalities through dynamically computed attention weights. We propose to compute the  $\mathbf{Z}_\text{MP}$ as the query, the $\mathbf{Z}_\text{CSI}$ as both key and value. This is because the query $\mathbf{Q}$ acts as a guide to determine which parts of the CSI feature are most relevant for channel extrapolation \cite{seneviratne2024cross}, while the feature of multi-path information $\mathbf{Z}_\text{MP}$ contains much rich information of the prorogation environment and is more beneficial for channel extrapolation. The above process is given mathematically as:
\begin{align}
\mathbf{Q} =\mathbf{W}_\text{q}\cdot\mathbf{Z}_\text{MP},\label{eq:Q}\\
\mathbf{K} =\mathbf{W}_\text{k}\cdot\mathbf{Z}_\text{CSI},\label{eq:K}\\
\mathbf{V} =\mathbf{W}_\text{v}\cdot\mathbf{Z}_\text{CSI},\label{eq:V}
\end{align}where $\mathbf{W}_\text{q},\mathbf{W}_\text{k},\mathbf{W}_\text{v}\in\mathbb{R}^{D\times D}$ are learnable weight matrices. These projected tensors are then split into $H$ heads and reshaped for parallel computation, can be denoted by $\mathbf{Q}_\text{h},\mathbf{K}_\text{h},\mathbf{V}_\text{h}\in\mathbb{R}^{B\times H\times L\times (D/H)}$. The attention weights are computed using the scaled dot-product attention mechanism:

\begin{equation}
\text{Attention}(\mathbf{Q}_\text{h},\mathbf{Q}_\text{h}) = \text{Softmax}(\frac{\mathbf{Q}_\text{h}\mathbf{Q}_\text{h}^\mathrm{T}}{\sqrt{D/H}}),
\end{equation}where the scaling factor $\sqrt{D/H}$ prevents excessively large dot-product values that can lead to gradient instability. The output of the multi-head attention is obtained by applying the attention weights to the value vectors, followed by concatenation and linear projection, expressed as:

\begin{equation}
\mathbf{X}_\text{fusion} = \text{Concat}(\text{Attention}(\mathbf{Q}_\text{h},\mathbf{K}_\text{h}), \mathbf{V}_\text{h})\cdot\mathbf{W}_\text{o},
\end{equation} where $\mathbf{W}_\text{o}$ is the output projection matrix.

\subsection{Decoder}
 The decoder is designed to be lightweight, featuring fewer layers and a smaller hidden dimension than the encoder to efficiently reconstruct the complete CSI. Notably, we introduce a residual connection between the feature fusion module and the decoder. As shown in Fig. \ref{fig:architecture}, the input the decoder, denoted as $\mathbf{X_\text{input}^\text{dec}}$, is formed by the element-wise addition of the feature fusion output and the masked CSI as:
 \begin{equation}
\mathbf{X_\text{input}^\text{dec}}=\mathbf{X_\text{fusion}\oplus}\mathbf{X^\text{masked}_\text{CSI}}.
 \end{equation}
 
 This residual connection effectively alleviates the gradient vanishing issue, improving stability and convergence speed. It also enriches the information flow to the decoder, boosting extrapolation accuracy. The decoding process begins by projecting the decoder input $\mathbf{X_\text{input}^\text{dec}}$ into a suitable dimensional space via a linear embedding layer:
 \begin{equation}
 \mathbf{Y}=\mathbf{X_\text{input}^\text{dec}}\mathbf{W}_\text{d}, 
 \end{equation} where $\mathbf{W}_\text{d}$ is a learnable projection matrix. 
 Same as the positional encoding of the encoder, channel-invariant 2D sine-cosine positional encoding $\mathbf{E}_\text{pos}^\text{dec}$ is added to incorporate spatial information:
 \begin{equation}
 \mathbf{Y}_\text{pos} = \mathbf{Y}+\mathbf{E}_\text{pos}^\text{dec}.
 \end{equation}
 
 The decoder applies a series of lightweight transformer blocks to process the sequence with restored structure. Each block contains a MSA module and a FFN interleaved with residual connections and LayerNorm. The computing of each transformer block for a given input $\mathbf{Y}_\text{pos}$ is given as:
 \begin{align}
 \mathbf{Y}_\text{block} = \text{LayerNorm}(\mathbf{Y}_\text{pos} + \text{MSA}(\mathbf{Y}_\text{pos})),\\ 
\text{MSA}(\mathbf{Y}_\text{pos}) = \text{Concat}(\text{head}_1, \dots, \text{head}_h)\mathbf{W}^O, 
 \end{align} where each head computes as: 
\begin{equation}
\text{head}_i = \text{Softmax}\left(\frac{(\mathbf{Y}_\text{pos}\mathbf{W}_i^\text{Q})(\mathbf{Y}_\text{pos}\mathbf{W}_i^\text{K})^\top}{\sqrt{d_k}}\right)\mathbf{Y}_\text{pos}\mathbf{W}_i^\text{V}. 
\end{equation}
The output is then computed as: 
\begin{align}
\mathbf{Y}_\text{dec} = \text{LayerNorm}(\mathbf{Y}_\text{block} + \text{FFN}(\mathbf{Y}_\text{block})),\\
\text{FFN}(\mathbf{Y}_\text{block}) = \text{GELU}(\mathbf{Y}_\text{block}\mathbf{W}_1 + \mathbf{b}_1)\mathbf{W}_2 + \mathbf{b}_2. 
\end{align}

After processing through all decoder blocks, the sequence is normalized by a final layer normalization:
\begin{equation}
\mathbf{Y}_\text{norm} = \text{LayerNorm}(\mathbf{Y}_\text{dec}). 
\end{equation} 

Finally, we map the latent representation back to the original space by using a linear projection and reshape it to the original CSI format:
\begin{equation}
\widehat{\boldsymbol{H}} = \text{Reshape}(\mathbf{Y}_\text{norm}\mathbf{W}_\text{r}).
\end{equation}

\section{Simulation Results}
\label{Sec:simulations}
\subsection{Simulation Settings}

\begin{figure}[]  
\centering
\includegraphics[width=0.5\textwidth]{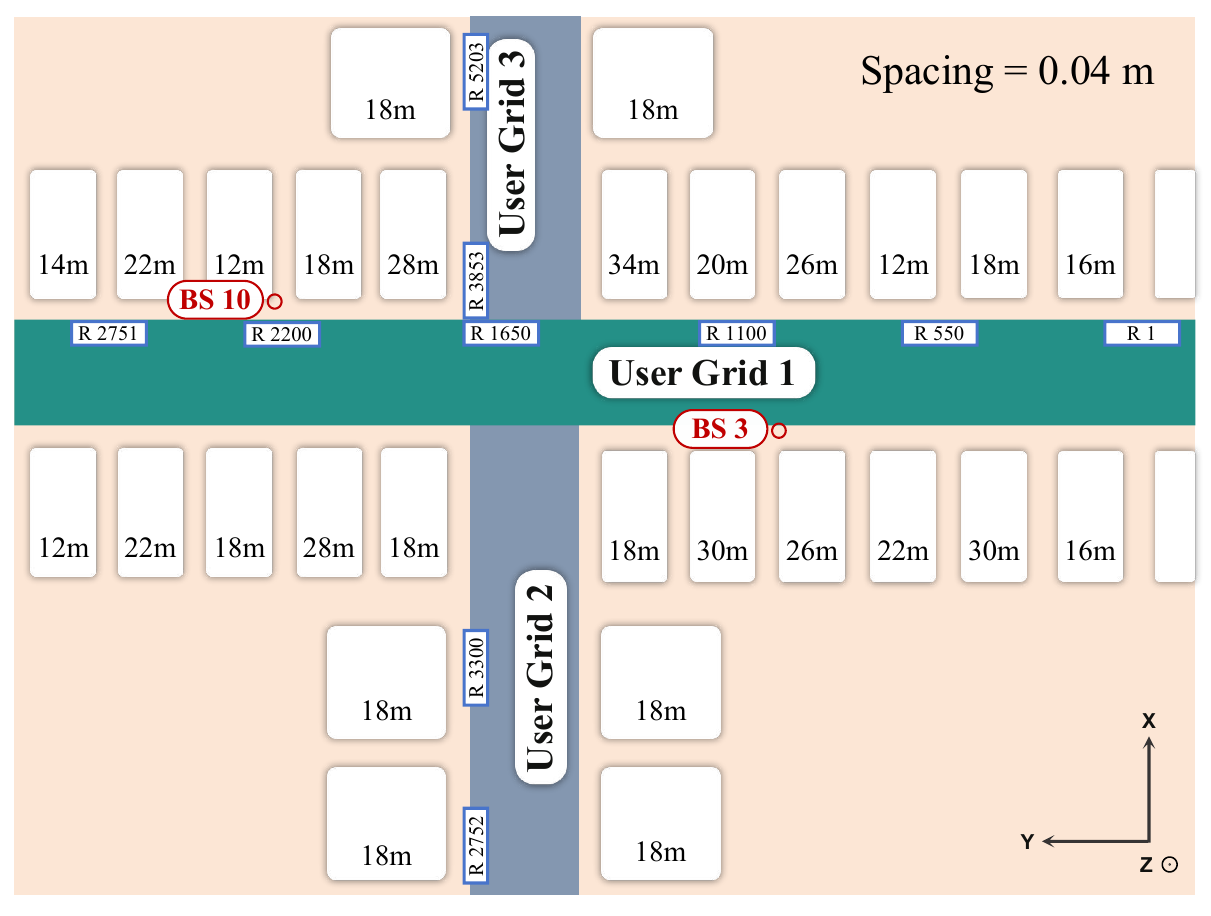} 
\caption{Simulation scenario.}
\label{fig:scenario}  
\end{figure}

\begin{table}[]
    \centering
        \caption{Simulation and training settings: \textbf{*} indicates settings for generalization evaluation. C2P and CE indicate CSI-to-PDP and channel extrapolation modules.}
    \label{table:simulation_settings}
    \begin{tabular}{cc}
    \toprule
        \textbf{Simulation parameters} & \textbf{Settings}\\\midrule 
        Scenario & O1 \\ 
        Active BSs & BS 3, BS 10 \\ 
        Active UEs & User Grid 1, 2 and 3 \\ 
        Number of BS antennas & 16 \\ 
        Number of UE antennas & 8 \\ 
        UE sampling factor & 0.1 \\ 
        {Center frequency} & 3.5 GHz, 28 GHz\textbf{*}\\ 
        Number of paths & 10 \\ 
        Number of subcarriers & 200 \\ 
        Bandwidth & 160 MHz \\ \midrule
        \textbf{Hyperparameter for training C2P} & \textbf{Settings} \\ \midrule
Hardware & NVIDIA GeForce RTX 4090\\
Batch size & 100\\
Optimizer & Adam\\
Learning rate & $1\times 10^{-4}$\\ \midrule
\textbf{Hyperparameter for training CE} & \textbf{Settings} \\ \midrule
Hardware & NVIDIA GeForce RTX 4090\\
Batch size & 100\\
Optimizer & AdamW\\
Learning rate & $1\times 10^{-3}$\\
Weight decay & 0.05\\
Betas & (0.9, 0.95)\\\bottomrule
\end{tabular}

\end{table}

\begin{figure}[t]
    \centering
    \subfigure[Training process of the CSI-to-PDP module.]{
        \includegraphics[width=\linewidth]{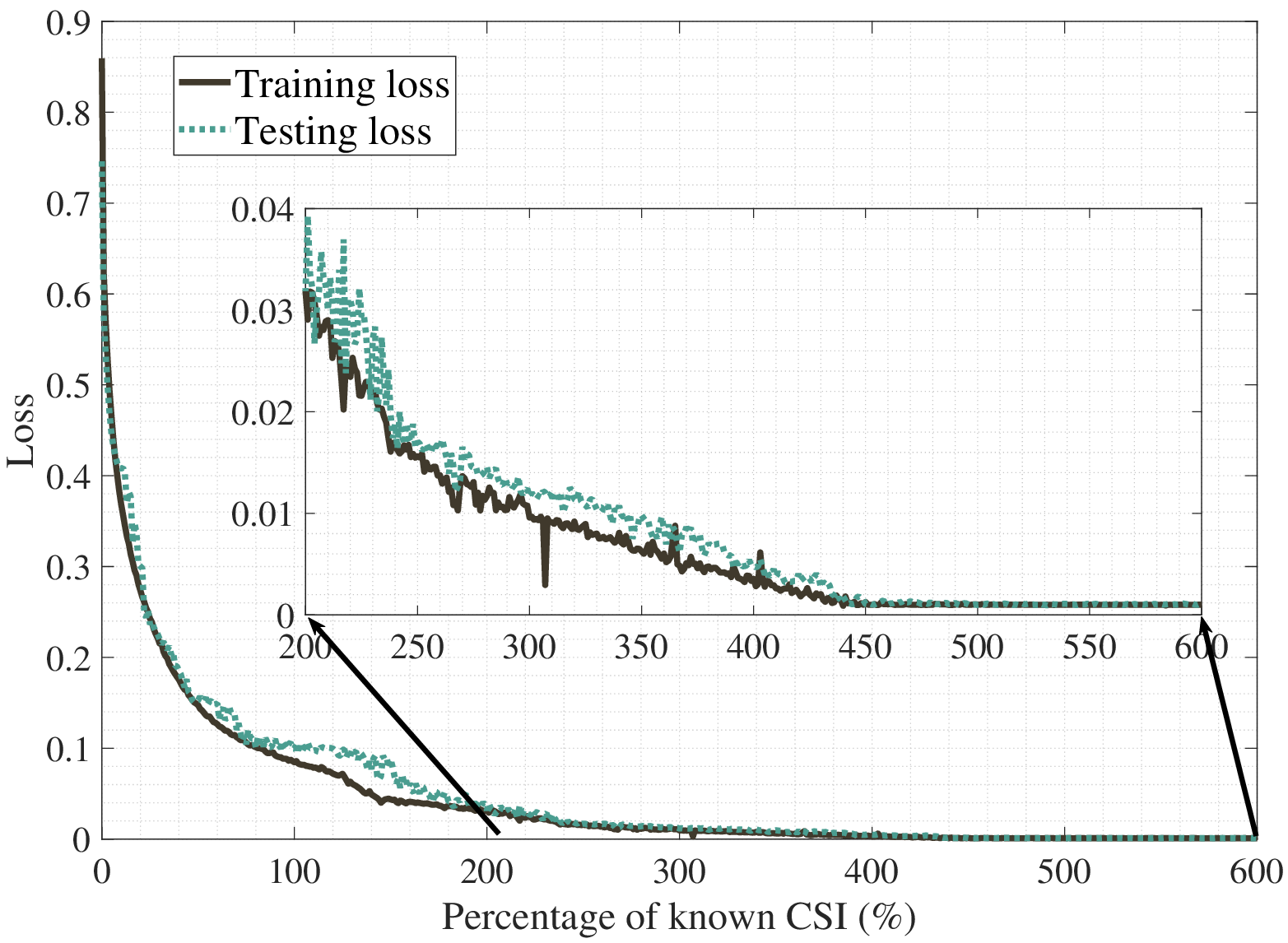}
        \label{fig:loss_AE}
    }
        \subfigure[Training process of the channel extrapolation module.]{
        \includegraphics[width=\linewidth]{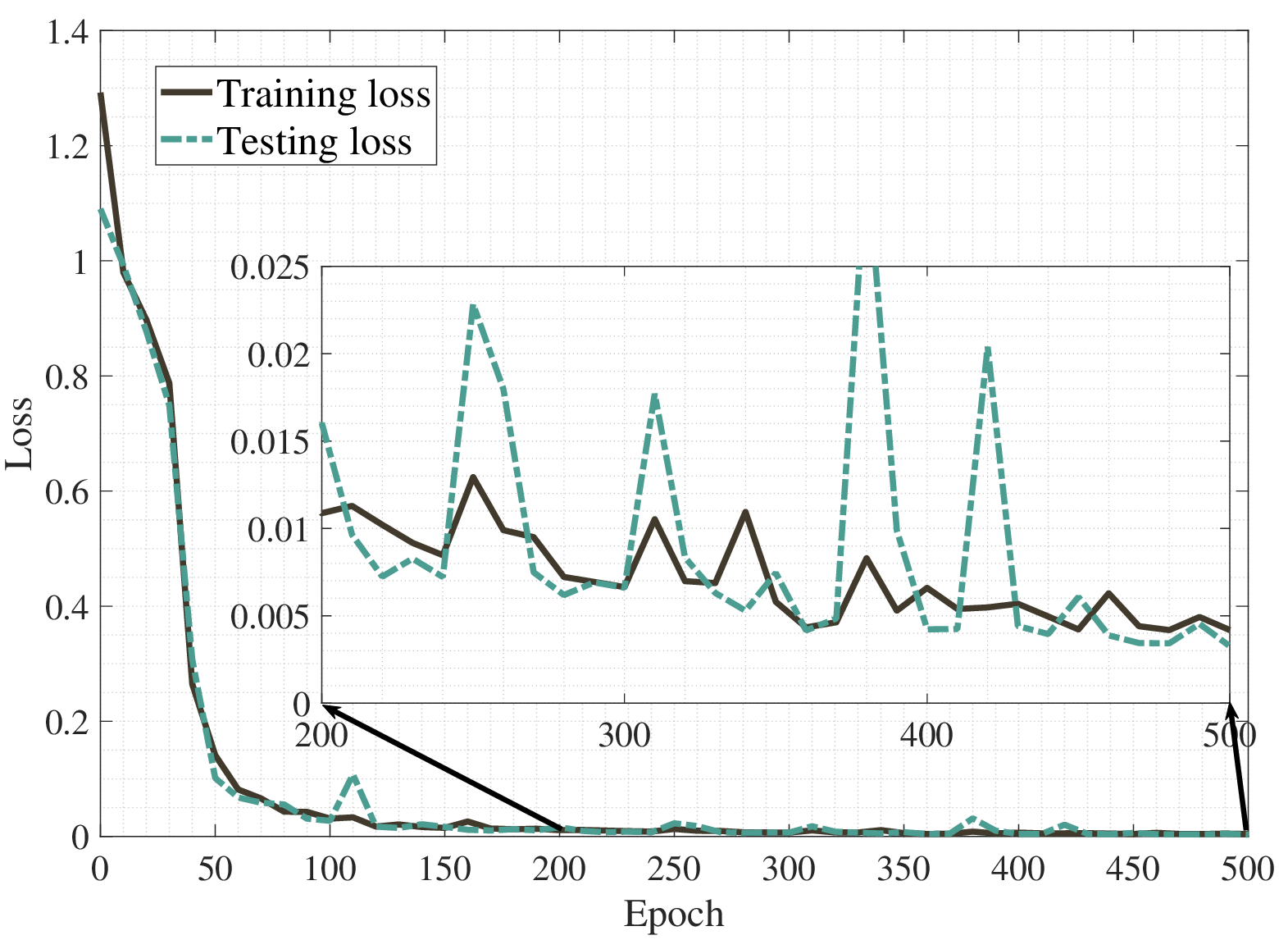}
        \label{fig:loss_MAE}
    }
    \caption{Illustration of the training and testing loss.}
    \label{fig:loss}
\end{figure}

\subsubsection{Data Generation}
In our experiments, we generate data by the DeepMIMO channel generation platform \cite{alkhateeb2019deepmimo}, which constructs MIMO channel data in different scenarios and parameter sets. As shown in Fig \ref{fig:scenario}, we select the scenario ‘O1’, an outdoor scenario consisting of two streets and one intersection. The main street (the horizontal one) is 600 m long and 40 m wide, and the second street (the vertical one) is 440 m long and 40 m wide. The experiment activates BS3 and BS10 as BSs, along with three user grids. We set the center frequencies to 5.9 GHz and 28 GHz. A total of 20,000 CSI samples are generated in 5.9 GHz at BS3, where 90 \% of these samples are used for training, and the remaining samples are used for validation. To evaluate the model’s generalization capability, an additional 5,000 CSI samples are produced in the same scenario with 28 GHz. Each CSI can be denoted as $\mathbf{h}_m\in\mathbb{C}^\mathbf{16\times8\times200}$ , where 16, 8 and 200 refer to the number of transmit antennas, receive antennas and subcarriers, respectively, with a subcarrier spacing set to 60 kHz. The other parameters are summarized in Table. \ref{table:simulation_settings}.

\subsubsection{Network Training}
The model was implemented using the PyTorch framework and trained on an NVIDIA GeForce RTX 4090 GPU. For the CSI-to-PDP network, we employed a batch size of 100 and utilized the Adam optimizer with a learning rate of $1\times 10^{-4}$. As shown in Fig. \ref{fig:loss_AE}, both the training and testing losses exhibit a consistent downward trend, ultimately converging after approximately 400 epochs. During the initial 100 epochs, the losses decrease sharply and rapidly, indicating the model's strong capacity for fast initial learning and its ability to promptly capture primary patterns and features from the training data. Beyond this phase, the rate of loss reduction becomes more gradual, reflecting that the model has moved past acquiring the most obvious features and has entered a phase of more intricate fine-tuning to minimize the loss further. In particular, training and testing losses remain closely aligned throughout the entire process, suggesting that the model generalizes well to unseen data without overfitting the training set.  

The channel extrapolation network utilizes the AdamW optimizer with an initial learning rate of $1\times 10^{-3}$, weight decay of 0.05, and betas set to (0.9, 0.95). The batch size is 100. The learning rate schedule with a 40-epoch warm-up phase followed by cosine annealing decay to a minimum of $1\times 10^{-6}$. In contrast to the smooth training of CSI-to-PDP in Fig. \ref{fig:loss_AE}, the training process of this module more complex. As shown in Fig. \ref{fig:loss_MAE}, both the training and testing losses show a declining trend and converge around 500 epochs. In the early stages of training, approximately within the first 150 epochs, the loss values undergo a rapid decline, demonstrating the model's strong initial fitting capability. However, after the loss decreases to a low level, the curves do not smoothly transition to a plateau as in Fig. \ref{fig:loss_AE}, but rather fluctuate around a low mean value, with the testing loss exhibiting particularly significant volatility, including several brief upward spikes. This oscillation convergence reflects that the model training process faces greater challenges, but the overall downward trend of the loss curve is maintained and eventually stabilizes in the ideal interval, indicating that the model still achieves effective convergence after overcoming the training challenges.

\subsection{Results Analysis}
%Fig. 1: loss convergence
%Fig. 3: multi-path characteristics accuracy

\begin{figure}[t]  
\centering
\includegraphics[width=0.5\textwidth]{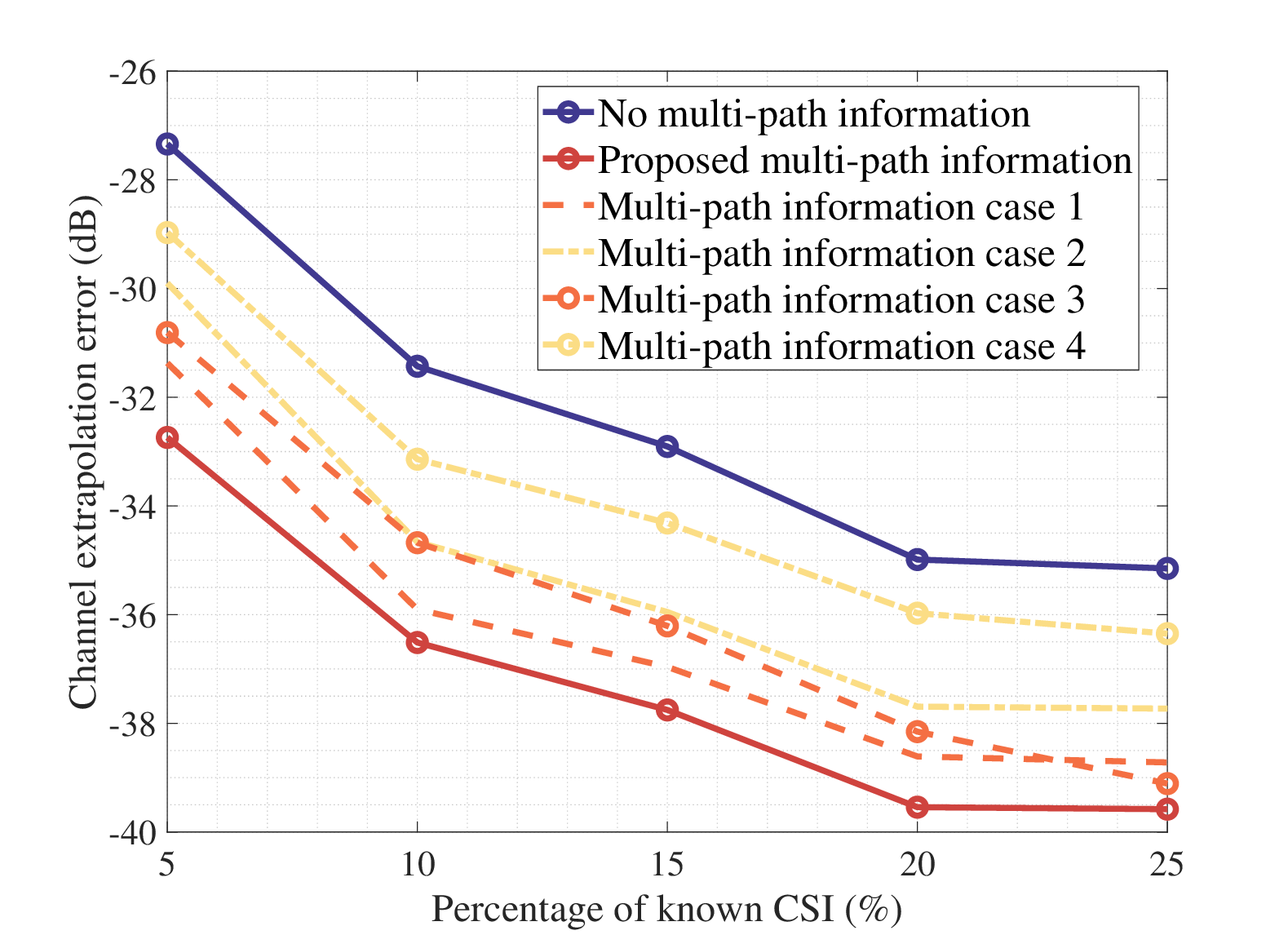} 
\caption{Comparison of the proposed with various type of multi-path information. Proposed multi-path information: sum of power and power-weighted delay (proposed); Case 1: power and delay of the first identified path; Case 2: power and delay of the path with largest power; Case 3: proposed \& Case 1; Case 4: proposed \& Case 2.}
\label{fig:comprison_multipath}  % 标签
\end{figure}

\subsubsection{Channel extrapolation performance}
Fig. \ref{fig:comprison_multipath} illustrates the channel extrapolation accuracy of the proposed model using different types of multi-path information and the baseline against various percentages of known CSI. 5 Types of multi-path information are considered, i.e., Proposed multi-path information: sum of power and power-weighted delay (proposed); Case 1: power and delay of the first identified path; Case 2: power and delay of the path with largest power; Case 3: proposed \& Case 2; Case 4: proposed \& Case 2. It is observed by using any type of multi-path information, the channel extrapolation accuracy is enhanced dramatically, which demonstrates the effectiveness of the proposed framework. Further, different types of multi-path information enhances the channel extrapolation with different extent. Specifically, proposed multi-path information, Case 1, Case 2, Case 3 and Case 4 outperforms the baseline with 5 \% of known CSI are 5.40, 4.03, 2.56, 3.47 and 1.63 dB, respectively. With a larger percentage of known CSI, the performance enhancement over the base line generally decrease, reaching 4.43, 3.57, 2.58, 3.96 and 1.20 dB for the proposed multi-path information, Case 1, Case 2, Case 3 and Case 4. Such observation demonstrates that the proposed model is more effective in the scenario its unique advantages. 

Another significant observation is that the proposed multi-path information is most effective to enhance the channel extrapolation. By comparing proposed multi-path information, case 1 and 2, the sum of power and power-weighted delay are more beneficial for channel extrapolation than power and delay of the first identified path, and power and delay of the path with largest power. This is attributed to that the sum of power and power-weighted delay contains much more comprehensive information of the propagation environment, thereby achieving the highest performance gain. Specifically, proposed multi-path information, which relies solely on the aggregate features (sum of power across paths and power-weighted delay), likely provides a more robust and concise representation of the overall propagation environment. This aggregation inherently smooths out path-specific variations, noise, or identification errors that could arise when estimating individual path parameters (e.g., due to limited known CSI). 

The proposed multi-path information outperforms Cases 3 and 4 is not intuitive to understand as Cases 3 and 4 consist of Case 1 and additional information. In contrast, Case 3 incorporates these aggregate features alongside the details of the first identified path (from Case 1), and Case 4 adds details from the strongest path (from Case 2). While intuitively additive information might seem beneficial, the specific path details could introduce correlated or redundant elements that overlap heavily with the aggregates, such as the first path often contributing disproportionately to the power sum or weighted delay in many scenarios, leading to feature redundancy. This redundancy can dilute the model's focus during training or inference, especially in low-data regimes like small percentages of known CSI, where the model is more prone to overfitting on noisy or inconsistent path-specific signals rather than leveraging the cleaner, holistic summary in the proposed multi-path information. As the proposed multi-path information has the best performance for the propose framework, we use the proposed multi-path information as the input for the remaining performance evaluation. 
\subsubsection{Feature fusion module}
\begin{figure}[t]  
\centering
\includegraphics[width=0.5\textwidth]{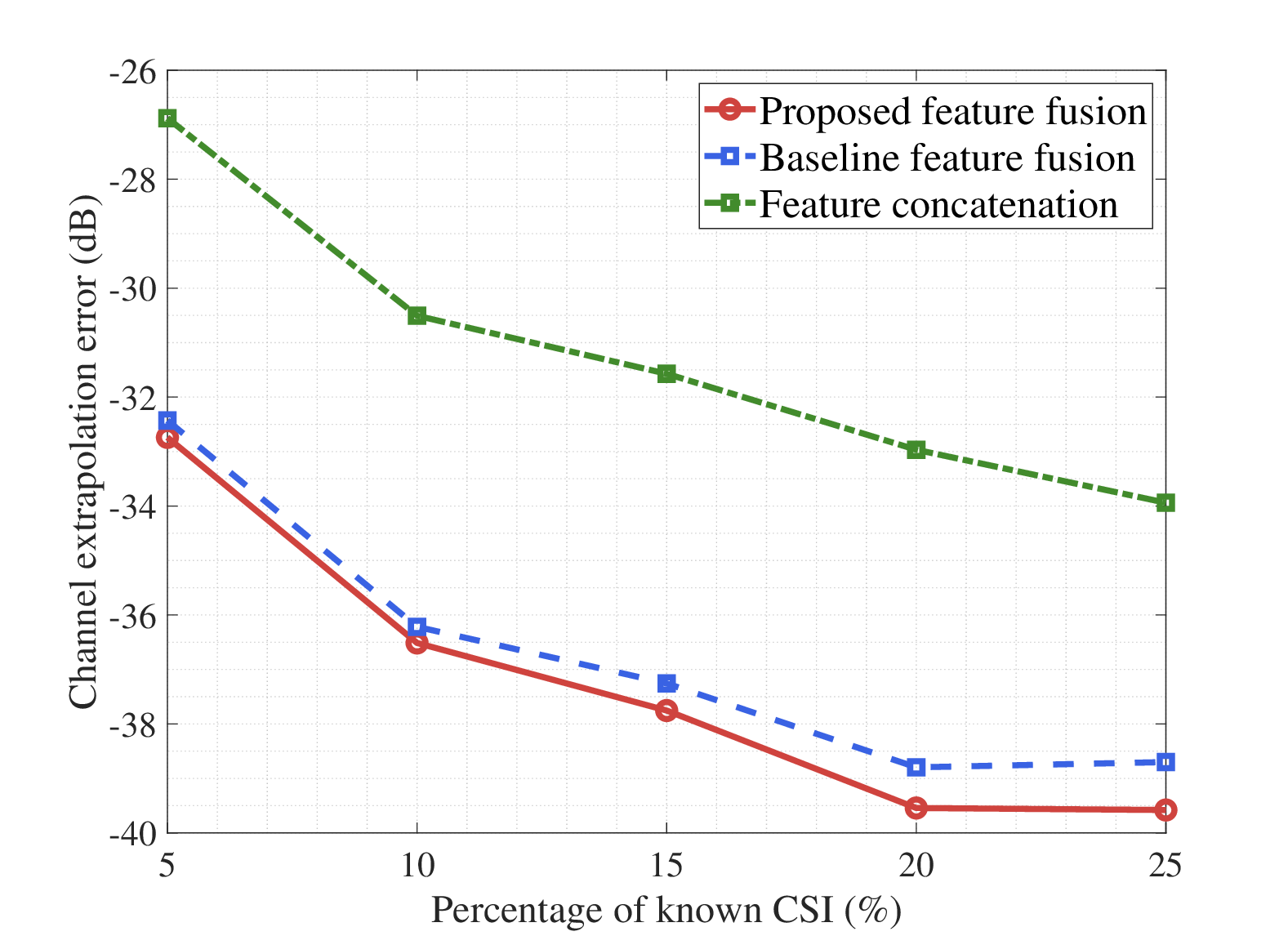} 
\caption{Comparison of the proposed with various types of fusion for CSI and multi-path information features. }
\label{fig:feature_fusion}  
\end{figure}
\begin{figure}[h]  
\centering
\includegraphics[width=0.5\textwidth]{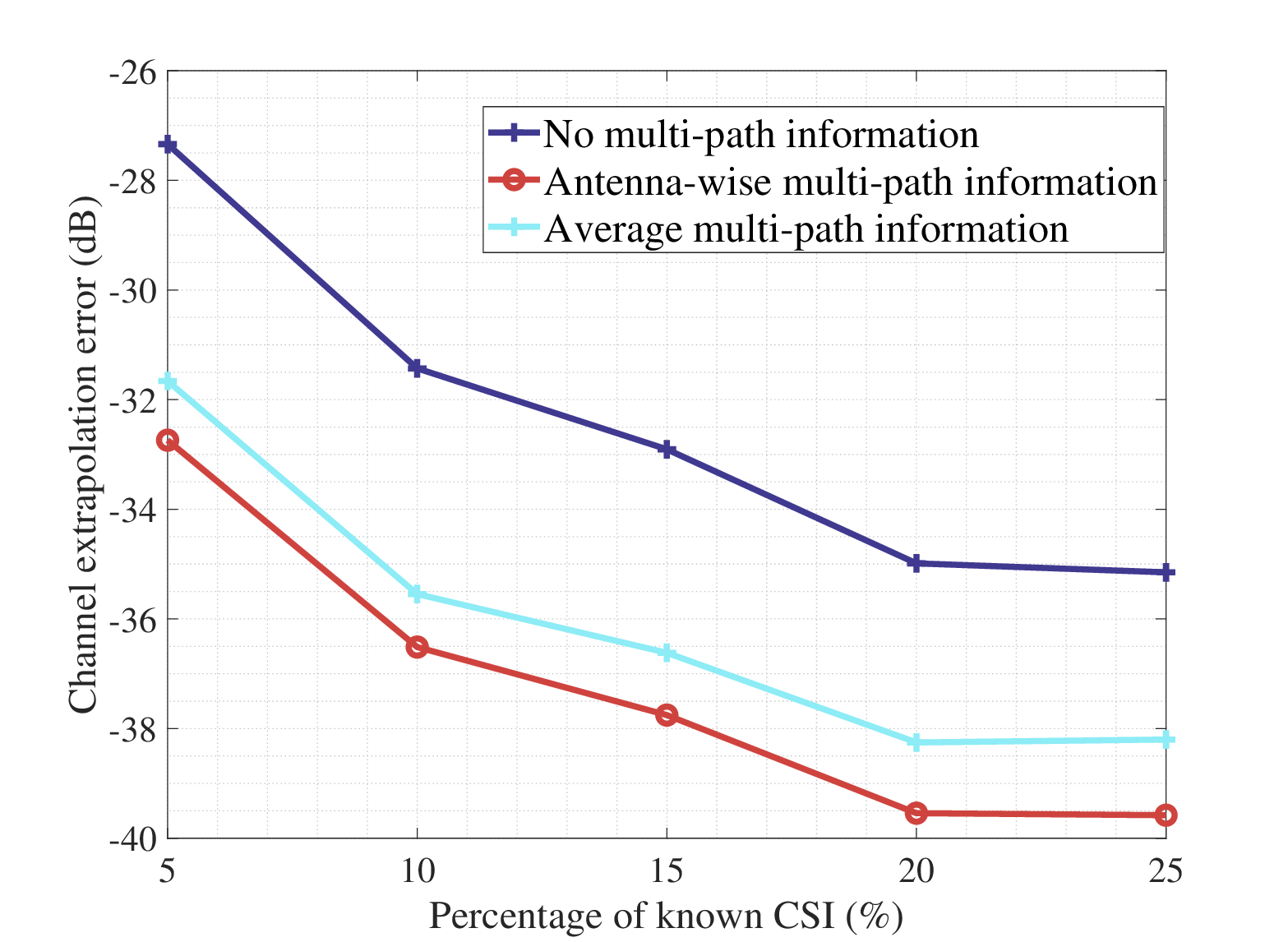} 
\caption{Comparison of antenna-wise multi-path information and average multi-path inforamtion.}
\label{fig:comprison_average_antenna_wise}  % 标签
\end{figure}

We demonstrate the effectiveness of the proposed fusion module of CSI and multi-path information features in Fig. \ref{fig:feature_fusion} by comparing with baseline feature fusion module via self-attention model and feature contamination. Different from the proposed feature fusion module given in (\ref{eq:Q})-(\ref{eq:V}), the baseline feature fusion module takes the feature of CSI $\mathbf{X}_\text{CSI}$ to compute query $\mathbf{Q}$, and the features of multi-path information $\mathbf{X}_\text{MP}$ to compute key $\mathbf{K}$ and value $\mathbf{V}$ as:
\begin{align}
\mathbf{Q} =\mathbf{W}_\text{q}\cdot\mathbf{X}_\text{CSI},\label{eq:Q1}\\
\mathbf{K} =\mathbf{W}_\text{k}\cdot\mathbf{X}_\text{MP},\label{eq:K1}\\
\mathbf{V} =\mathbf{W}_\text{v}\cdot\mathbf{X}_\text{MP}.\label{eq:V1}
\end{align}

The feature contamination module simply contaminates feature of CSI $\mathbf{X}_\text{CSI}$ and features of multi-path information $\mathbf{X}_\text{MP}$ as $\text{Concat}(\mathbf{X}_\text{CSI},\mathbf{X}_\text{MP})$. 

As illustrated in Fig. \ref{fig:feature_fusion} the proposed feature module outperforms the scheme of feature concatenation dramatically by about 6 dB, indicating that self attention module is more effective to exploit the feature of CSI and multi-path information. The comparison between the proposed feature fusion module and the baseline feature fusion module via self-attention model indicates that the choice of features to compute query $\mathbf{Q}$, key $\mathbf{K}$ and value $\mathbf{V}$ makes difference. Specifically, using multi-path information $\mathbf{X}_\text{MP}$ to compute query $\mathbf{Q}$, and $\mathbf{X}_\text{CSI}$ to compute key $\mathbf{K}$ and value $\mathbf{V}$ is more effective than using multi-path information $\mathbf{X}_\text{MP}$ to compute key $\mathbf{K}$ and value $\mathbf{V}$, and $\mathbf{X}_\text{CSI}$ to compute query $\mathbf{Q}$. This is because the feature of multi-path information $\mathbf{X}_\text{MP}$ contains much rich information of the prorogation environment and is more beneficial for channel extrapolation, while the query $\mathbf{Q}$ acts as a guide to determine which parts of the CSI feature are most relevant for channel extrapolation \cite{seneviratne2024cross}. The baseline uses CSI feature $\mathbf{X}_\text{CSI}$ to compute query $\mathbf{Q}$, which lacks the same environmental richness, leading to less effective attention weighting. This results in suboptimal fusion, as the model struggles to identify which multi-path information are most relevant for extrapolating unknown CSI.

\begin{table*}[h]\centering\caption{Comparison of the baseline and the proposed framework in terms of inference speed.}\label{table:inference_time}
\begin{tabular}{ccccc}
\toprule
                & Baseline (ms) & Proposed framework (ms) & Delay increase (\%) & Delay increase (ms) \\ \midrule
5 \% known CSI  & 0.97          & 1.12                    & 15.1                & 0.15                \\ 
10 \% known CSI & 1.01          & 1.17                    & 15.8                & 0.16                \\ 
15 \% known CSI & 1.12          & 1.21                    & 8.0                 & 0.09                \\ 
20 \% known CSI & 1.17          & 1.26                    & 7.7                 & 0.09                \\ 
25 \% known CSI & 1.20          & 1.31                    & 9.2                 & 0.11                \\ \bottomrule
\end{tabular}
\end{table*}

\subsubsection{Type of environment information}

\begin{figure}[]
    \centering
    \subfigure[The power]{
        \includegraphics[width=\linewidth]{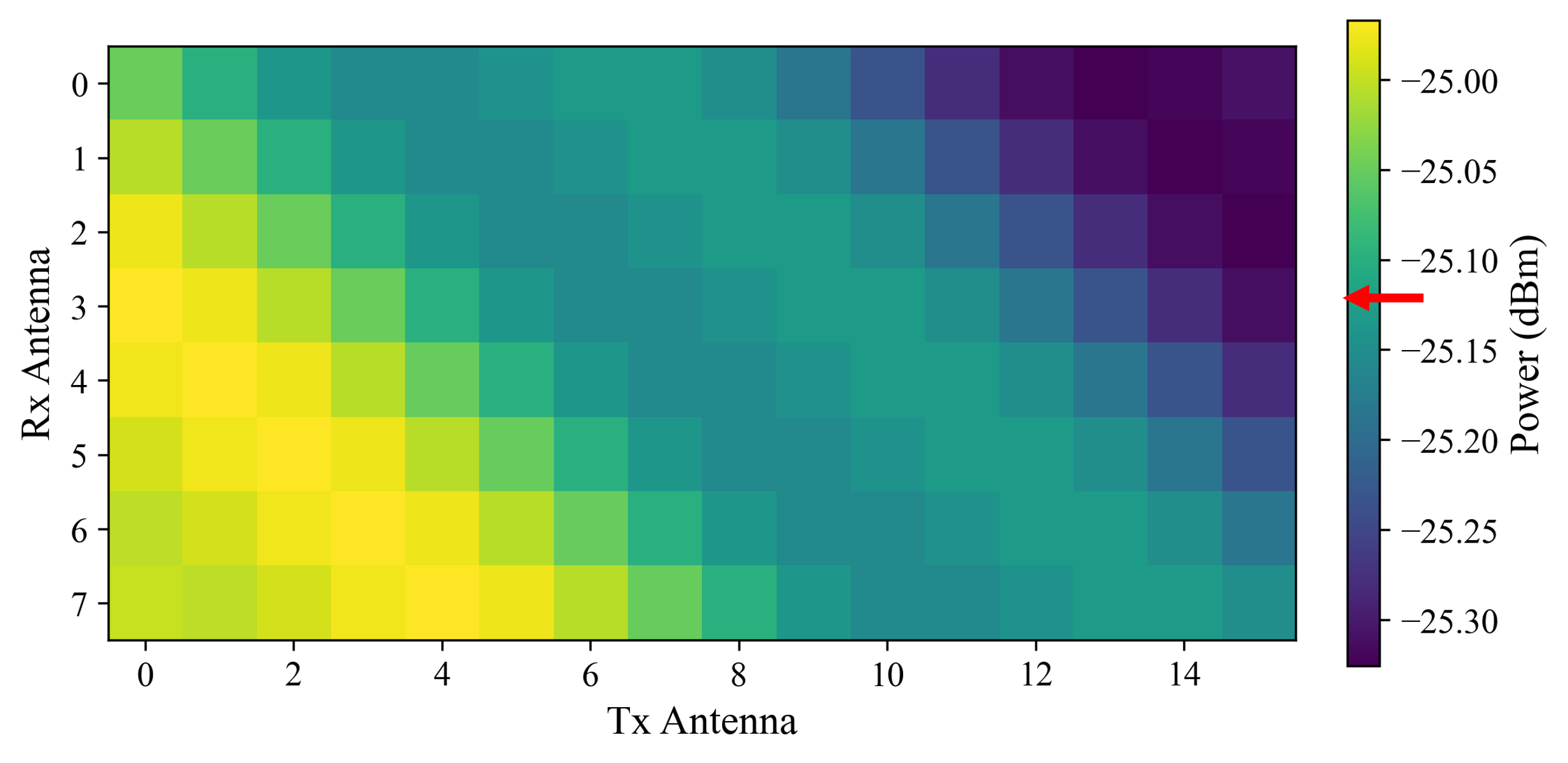}
        \label{fig:power}
    }
        \subfigure[The power-weighted delay.]{
        \includegraphics[width=\linewidth]{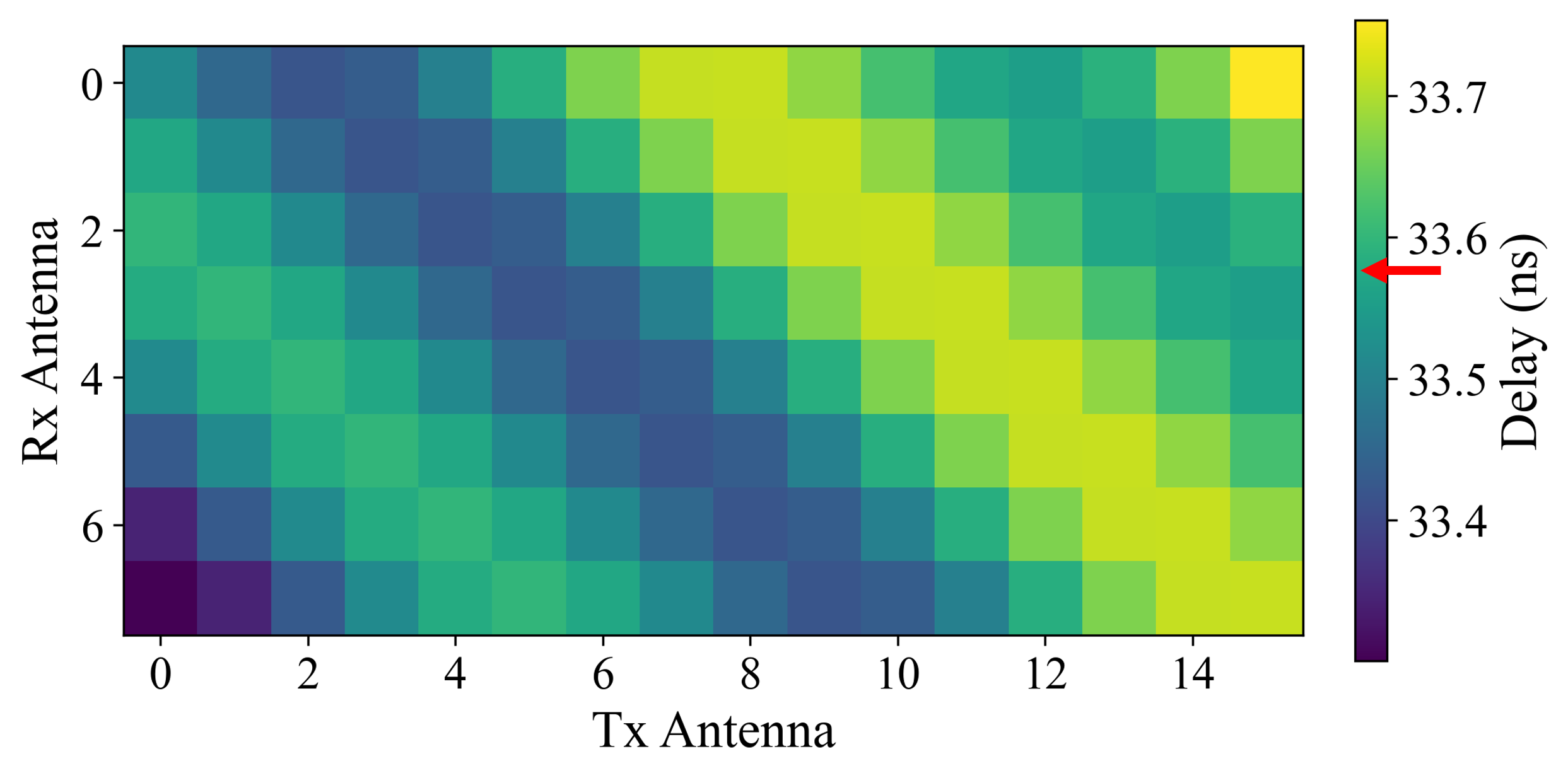}
        \label{fig:weighted-delay}
    }
    \caption{Illustration of the antenna-wise and average multi-path information. The red arrows denote the average values of power and power-weighted delay.}
    \label{fig:average_antenna_wise}
\end{figure}
\begin{figure}[]
    \centering
        \includegraphics[width=1\linewidth]{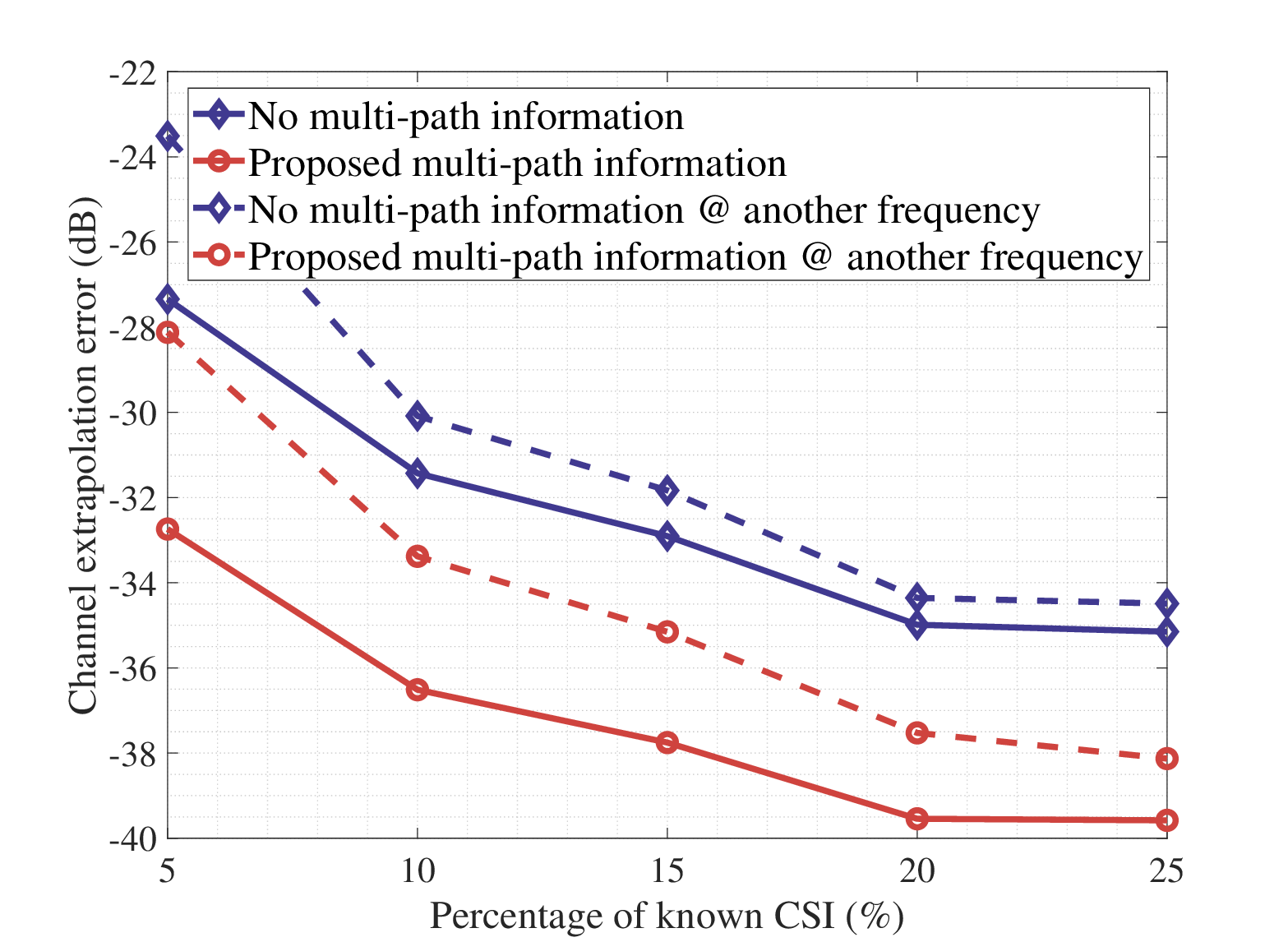}
    \caption{Generalization performance of the proposed model tested at another frequency.}
    \label{fig:frequency}
\end{figure}

In the proposed framework, the multi-path information of all the observable antennas are used as input. A question arises that whether antenna-wise multi-path is necessary? Can we use the average multi-path information across the antennas as the input to strike a balance between the computational complexity and performance? To shed light on the above two questions, we first illustrate the the antenna-wise power and power-weighted delay for the MIMO systems of interest in Figs. \ref{fig:power} and \ref{fig:weighted-delay}, which demonstrates a specific pattern of the power and power-weighted delay across antennas. It is noted that range of power and power-weighted delay is quite small, which are 1.5 and 1.6 \%, respectively. Fig. \ref{fig:comprison_average_antenna_wise} illustrates that the proposed framework with antenna-wise multi-path information outperforms that with the average multi-path information by 1.12 and 1.51 dB with 5 \% and 25 \% of known CSI, respectively. Such notable gain indicates that the pattern of the antenna-wise multi-path information is beneficial for channel extrapolation.

\subsubsection{Generalization performance}
As a critical aspect for the applicability in practical wireless, the generalization of the proposed framework is evaluated in a zero-shot leaning manner in different carrier frequency in Figs. \ref{fig:frequency}, which illustrates the channel extrapolation performance of the proposed framework and the baseline model evaluated in 28 GHz scenarios. The proposed framework outperforms the baseline model by 4.61 and 3.64 dB with 5 \% and 25 \% of known CSI, respectively. Such notable performance enhancement is attributed to the fact that the directions of the multi-path in 28 GHz (test scenario) are pretty much aligned with those of the multi-path in 5.9 GHz (training scenario), but with much lower power amplitude due to the dramatic propagation attenuation of 28 GHz compared to that of 5.9 GHz \cite{zhang2023ai}.

%Different from the generalization performance in 20 GHz, the generalization performance of the proposed framework with different BS locations is much less significant. As illustrated in Fig. \ref{fig:generalization_location}, the proposed framework outperforms the baseline model by 1.71 dB with 5 \% of known CSI, and tend to vanish with the increase of the percentage of known CSI. This is attributed that the shared information of multi-path information in different locations are much less, leading to less performance enhancement. However, the performance enhancement with low percentage of known CSI is still achieved, which is inspiring to reduce the overhead. 

\subsubsection{Inference speed}
The additional input and processing of the multi-path information in the proposed framework can effectively enhance the performance of the channel extrapolation, and inevitable increase its computational complexity. Table. \ref{table:inference_time} illustrates the inference time of the proposed framework and the baseline with various percentages of known CSI. The increase of the inference time is at the level of 0.1 ms and around 10 \%, which is acceptable. This demonstrates that a slight increase of the computational complexity can effectively enhance the channel extrapolation performance.

\section{Conclusions}
\label{Sec:conclusions}
This paper has proposed enhancing channel extrapolation by leveraging environmental information without integrating additional modalities. The environmental information in the form of PDP is acquired directly using the known CSI by an AE-based framework. We further proposed dedicated encoder to extract to features of CSI and multi-path information (extracted from PDP), which are further fused using a self-attention module to take the maximum advantages of the environmental information. Extensive simulations have demonstrated that this framework improves extrapolation performance by approximately 4–5 dB, with a minor increase in inference time (around 0.1 ms) on an NVIDIA GeForce RTX 4090 GPU. Notably, the most influential multi-path information for high-performance channel extrapolation are the total path power and the power-weighted delay, which encapsulate comprehensive environmental information. Furthermore, our model has shown strong generalization capabilities across different frequency bands, particularly when only a small portion of the CSI is known, outperforming existing benchmarks.

\ifCLASSOPTIONcaptionsoff
  \newpage
\fi
 \small
\nocite{*}
\bibliography{reference.bib}
\end{document}